\documentclass[%
 reprint,
 amsmath,amssymb,
 aps,
]{revtex4-2}

\usepackage{graphicx}
\usepackage{dcolumn}
\usepackage{bm}
\usepackage{soul}
\usepackage{multirow}%
\usepackage{amsmath,amssymb,amsfonts}%
\usepackage{amsthm}%
\usepackage{mathrsfs}%
\usepackage{xcolor}%
\usepackage{float}
\usepackage{siunitx}

\begin{document}

\preprint{APS/123-QED}

\title{Driven Dissipation and Stabilization of a Levitated Rotor with Multi-Day Coherence}

\author{Mehrdad M. Sourki$^1$}
\author{Wisdom Boinde$^1$}
\author{Gautham Anne$^2$}
\author{Mahdi Hosseini$^{1,3,*}$}
 \affiliation{$^1$ Department of Electrical and Computer Engineering, Northwestern University, 2145 Sheridan Rd, Evanston, 60028, IL U.S.A}
 \affiliation{$^2$ Department of Mechanical Engineering, Northwestern University, 2145 Sheridan Rd, Evanston, 60028, IL USA}
 \affiliation{$^3$ Applied Physics Program, Northwestern University, 2145 Sheridan Rd, Evanston, 60028, IL U.S.A}
\email{mh@northwestern.edu}
\date{\today}

\begin{abstract}
We report the experimental observation of nonlinear mode-crossing dissipation and injection-locking–like stabilization in a milligram-scale diamagnetically levitated quartz cube. By optically driving the cube to rotation rate near 360 RPM, we observe that its angular velocity does not decay smoothly but instead decreases in discrete steps whenever the instantaneous rotation frequency crosses one of several higher-order mechanical modes. We observe rotational dissipation times on the order of a few days. When a weak counteracting radiation-pressure torque is applied, the system unexpectedly stabilizes at a constant rotation rate that persists for about two days—constituting the mechanical analogue of injection locking. Our results reveal a new regime of multimode nonlinear dynamics in low-dissipation levitated solids and establish diamagnetic levitation as a platform for exploring non-Hamiltonian many-mode interactions and mechanically engineered nonlinear attractors and sensors.
\end{abstract}

\maketitle


\section*{Introduction}\label{sec1}


Levitated mechanical systems provide a powerful platform for exploring linear and nonlinear  dynamics with exceptionally low dissipation and high environmental isolation. Recent advances in optical, magnetic, and diamagnetic levitation have enabled long coherence times, precise control of translational and rotational degrees of freedom, and applications ranging from force and inertial sensing to tests of fundamental physics \cite{Millen2020Review, Moore2021Review,kilian2024dark, higgins2024maglev, gonzalez2021levitodynamics}. At the same time, these systems naturally access regimes where linear intuition fails: weak damping, strong mode coupling, and external driving can give rise to nonlinear responses, mode hybridization, and emergent steady states \cite{Aspelmeyer2014RMP, RevModPhys.65.851}. Understanding how such nonlinear interactions redistribute energy and stabilize motion is therefore essential both for practical device operation and for uncovering general principles of non-equilibrium dynamics in mesoscopic mechanical systems.

In many nonlinear systems far from equilibrium, persistent and coherent motion emerges despite dissipation through the interplay of nonlinearity, continuous energy injection, and mode coupling. Prominent examples range from phase locking and injection locking in nonlinear oscillators \cite{Adler1946, Pikovsky2002} to long-lived coherent structures in fluid dynamics, such as Jupiter’s Great Red Spot, which is stabilized by nonlinear interactions within a driven, dissipative environment \cite{Ingersoll1969, Marcus1993}. In mechanical systems, the experimental observation of mechanical frequency combs ~\cite{mechanicalComb2025,de2023mechanical}, nonlinear mode hybridization~\cite{matheny2014phase},
parametric amplification ~\cite{PhysRevLett.67.699, PhysRevLett.110.127202}, Fermi-Pasta-Ulam-Tsingou Recurrences ~\cite{sourki2025nonlinear}, and a host of effects have been observed that mirror well-established phenomena in optics but arise from entirely different microscopic origins.
Despite this rapid growth, the nonlinear dynamics of high-$Q$, macroscopic mechanical oscillators remains far less understood than its optical counterparts. Mechanical systems often possess large hierarchies of weakly coupled modes, highly nonlinear geometric interactions, and complex forms of dissipation, which together generate behaviors that do not obey simple perturbative theories. 

A particularly promising class of systems for studying nonlinear dynamics and developing  sensors includes diamagnetically levitated mechanical oscillators \cite{kim2025magnetically, matsko_mechanical_2020, chen2025levitatedmacroscopicrotors10, leng_mechanical_2021, sourki2025nonlinear}, where dissipation can be reduced to extraordinary levels and nonlinearity can be enhanced by field curvature, geometric asymmetry, or radiation-pressure driving. In our recent work, milligram-scale diamagnetically levitated quartz cubes~\cite{sourki2025nonlinear} have been shown to enter a regime of  coherent multi-mode nonlinear dynamics in which translational, torsional, and mixed rotational-translational modes interact over extremely long timescales. These systems provide an opportunity to explore dynamical phenomena that have no direct counterpart in linear mechanics, including nonlinear mode crossings, discrete dissipation bursts, and emergent attractors.

\begin{figure*}
    \centering
    \includegraphics[width=0.8\linewidth]{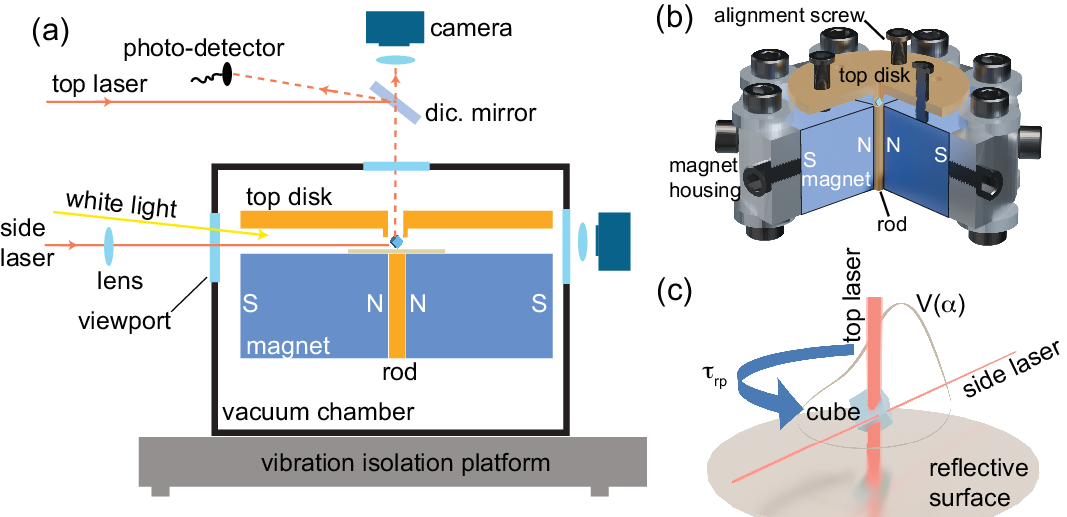}
\caption{(a) Schematic of the experimental platform. A continuous-wave 794 nm laser is used to illuminate the cube from the side and top. Two imaging systems—a top-view camera aligned with the vertical axis and a side-view camera positioned orthogonally—allow simultaneous measurement of translational and rotational motion with sub-micrometer precision. A photo-detector recording the reflection  of the top laser also measures the shadow of the cube to create the vibrational spectrum. The diamagnetic trap and optical access are mounted on an actively damped vibration-isolation stage inside an acoustic enclosure.  The vacuum enclosure has multiple windows on the side and at the top, enabling optical access for both driving and detection. (b) A section-view magneto-gravitational potential generated by an array of triangular wedge rare-earth permanent magnets. A top disk and a center rod made from permendur are used to focus the magnetic field. The levitation height and the trap axis is adjusted by a set of three alignment screws controlling the angle and position of the top disk.  (c) Schematic of radiation pressure torque ($\tau_{rp}$) applied to the cube. A weak radiation-pressure torque applied from the top laser and a strong torque from the side laser. The magnetic potential possesses a shallow angular barrier ($V(\alpha)$) near the equilibrium orientation, producing a natural “rocking” mode with frequency $f \sim 1-2~Hz$. When the laser is incident on the cube, the resulting torque can provide enough energy to push the system over this barrier, initiating sustained rotation.}
    \label{fig1}
\end{figure*}

Here, we explore driven rotation in the system in the presence of nonlinear mode coupling. When the cube is illuminated from the side using a focused off center laser, radiation pressure exerts a torque that spins the body to more than 360 RPM. Because of the object’s extreme isolation—its quality factors exceed those of most macroscopic oscillators—the rotational mode does not simply decay exponentially when the torque is removed. Instead, we observe a discrete staircase of frequency drops. As the rotation slows, its frequency sweeps through the higher order vibrational and rotational modes of the levitated cube. Each crossing leads to a burst of hybridization, a rapid loss of angular momentum, and a step-like change in rotation rate. The cube sheds rotational energy into fast-decaying internal motions through nonlinear couplings, much as we previously observed in the translational and torsional mode structure of this system \cite{sourki2025nonlinear}.

The presence of these discrete decay steps already suggests an analogy to mode hopping in multimode lasers, where the cavity selects one resonant branch until a drift in pump conditions forces a jump to another. But the analogy becomes much stronger—and more surprising—when the system is weakly driven in reverse. If a small counteracting radiation pressure is applied while the cube rotates, the torque slows the motion, arrests it, reverses its sign, and eventually induces rotation in the opposite direction. During this upward sweep in frequency we again observe mode crossings and discrete steps. Yet, remarkably, the cube then enters a regime in which the rotation locks to a single, stable frequency and remains fixed for hours to days. Despite the existence of nonlinear couplings and many opportunities to transfer energy into lossy internal modes, the rotation becomes immune to further discrete jumps. The dynamics settle into a near-steady state, not because dissipation has been eliminated, but because the weak external drive continuously nudges the system onto a stable attractor.

This phenomenon is strikingly reminiscent of injection locking in lasers. There, an oscillator with many unstable or competing modes can be coerced into a single well-defined frequency by a weak external signal. The injected tone stabilizes what would otherwise be a wandering system, suppressing mode hops and binding the oscillator to an externally defined frequency. In our levitated cube, radiation pressure plays the role of the injected tone. The nonlinear couplings between rotation and the cube’s translational mechanical modes create the equivalent of a multimode gain medium. When the rotation frequency approaches a nonlinear resonance, rather than producing a mode hop, the torque provides just enough phase and amplitude correction to hold the system to a single frequency. To our knowledge, this is the first observation of injection-locking-like stabilization in a macroscopic mechanical object with no contact-based dissipation.

In our system, a diamagnetically levitated cube exhibits a mechanical counterpart of this behavior: nonlinear coupling among a finite set of rotational and translational modes, when combined with even weak external driving, stabilizes persistent rotation and phase locking in a regime where linear theory would predict continuous decay or mode hopping. The system occupies an intermediate regime between low-dimensional chaotic models and high-dimensional turbulent systems, allowing emergent stabilization to be observed at the level of individual mechanical modes.

Taken together, these observations place the levitated dielectrics of asymmetric shapes in a new category of driven nonlinear mechanical systems—one that combines high coherence, strong mode mixing, and externally induced frequency locking.  

\section*{Experiment}
 A quartz cube of 0.5 mm size is levitated on a specifically designed arrangement of magnets allowing stable levitation of such a macroscopic object with no external power and with ultra-low dissipation rates. The levitation is performed under a vacuum pressure of $5\times10^{-8}$ Torr. Fig.~\ref{fig1} (a) shows a schematic of the experimental setup. A Ti:sapphire laser (M Squared) is used both to drive the cube and to probe its motion. A portion of the beam (the pump), used to drive rotation, is injected into the vacuum chamber through a lateral optical access. Similar to our previous work ~\cite{sourki2025nonlinear}, levitation is enabled by an array of triangular wedge magnets. A section-view of the magnet assembly is shown in Fig.~\ref{fig1} (b). 
 
 The chamber provides multiple viewports, which we exploit to introduce white-light illumination and to image the levitated cube from the side with a camera. The probe beam (top laser) is injected through the top optical access. A reflective disk placed beneath the levitated cube retro-reflects the probe; as illustrated in Fig.~\ref{fig1} (c), the probe and its reflection pass through the cube and acquire motion-dependent modulation. The reflected light is collected on a single-pixel photodetector. In addition, a dichroic mirror enables simultaneous top-view imaging of the levitated cube.  

\section*{Optical excitation through radiation pressure}
In steady state without an external drive, the cube is stably trapped in all directions including the rotation around the vertical axis (angle $\alpha$), as shown in Fig.~\ref{fig1} (c). This is because of the asymmetry in the shape of the levitated object and also rotational asymmetry of the magnetic field in the trapping region.

\begin{figure}
    \centering
    \includegraphics[width=1\linewidth]{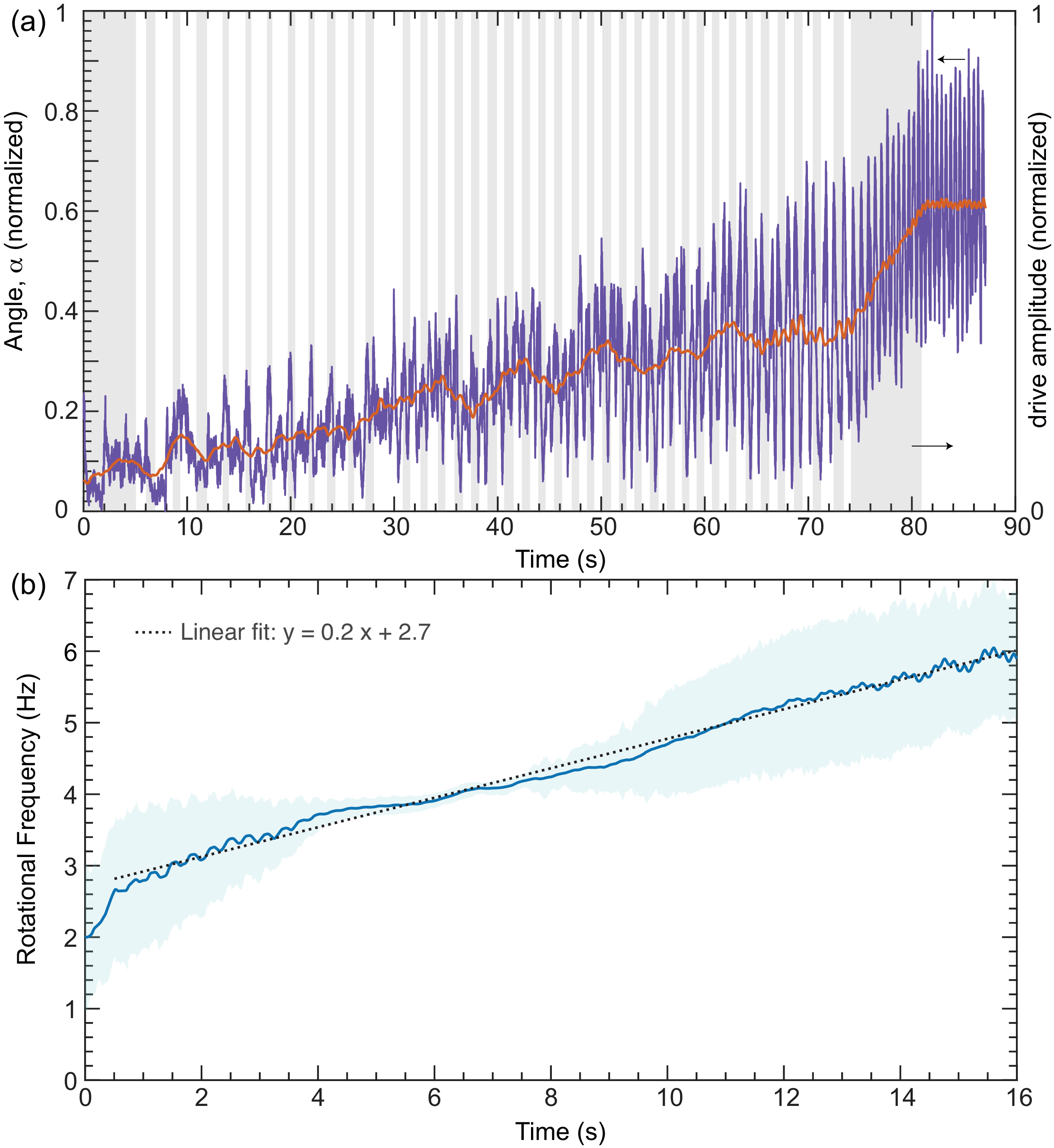}
    \caption{(a) Time-resolved video analysis of the angular dynamics of a levitated cube subjected to modulated radiation pressure. The blue trace shows the instantaneous angular excursion extracted from top-view imaging, while the red curve represents the same signal after temporal smoothing to reveal the underlying evolution. Gray shaded regions indicate intervals during which the radiation-pressure laser is on (right axis). Initially, the motion is confined to bounded rocking in an effective angular potential. Repeated radiation-pressure excitation leads to a gradual increase in the rocking amplitude until the particle overcomes the angular barrier and transitions into sustained rotation. (b) Rotation frequency (extracted from top-view video by tracking the angular position of a surface feature) as a function of time, during which a continuous radiation pressure torque was applied to the cube. The frequency is obtained from coarse-grained differentiation of the unwrapped angle. Shaded region indicates a $95\%$ confidence band estimated from local residual scatter.}
    \label{fig2}
\end{figure}

When the particle is mechanically excited (e.g. by applying an impulsive force to the table) different modes can be excited. The rotational mode around the vertical axis is seen as rocking motion at a frequency of $\sim \SI{1}{Hz}$. For a perfect trap, it is expected that rotational symmetry would make the rotational mode unhindered, but in the experiment, the trap imperfection and the directional magnetic susceptibility of quartz causes the cube to experience a potential barrier ($V(\alpha)$) in its rotational motion. Given the moment of inertia for the cube and the observed rocking motion frequency, we calculate the height of this barrier to be on the order of $\sim 10^{-14}\si{J}$. To overcome this barrier and selectively excite the rotational mode, we devise an experiment to drive the rotational modes optically relying on the radiation pressure force. This can be done using a focused laser from the top or side with optical power between $\SIrange{1}{100}{mw}$ (depending on the desired acceleration and the beam position). We first modulate the laser power driving the cube from the side nearly in-sync with the rocking motion. The result of angular amplitude over time is shown in Fig. \ref{fig2} (a), where clear excitation can be observed. Beyond a certain drive amplitude, the rocking motion overcomes the rotational potential barrier and the cube starts to spin. Fig. \ref{fig2} (b) shows that in this regime where about $\SI{100}{mW}$ optical power is continuously applied, the cube's rotational frequency almost linearly increases with time up to some point. Beyond a certain rotational energy the coupling to other modes leads to chaotic behavior and thus loss of the particle from the trap. Example videos illustrating pulsed and continuous optical driving are provided as Supplementary Videos 1 and 2, respectively.

The radiation pressure force depends on the power, surface reflection and the angle of incidence. The linear rotational frequency increase over time is a clear indication of radiation pressure force. We also observe the sensitivity of exerted torque to beam alignment, although it is difficult to quantify the angle and reflection given the size of the cube.  

\begin{figure*}
    \centering
\includegraphics[width=0.95\linewidth]{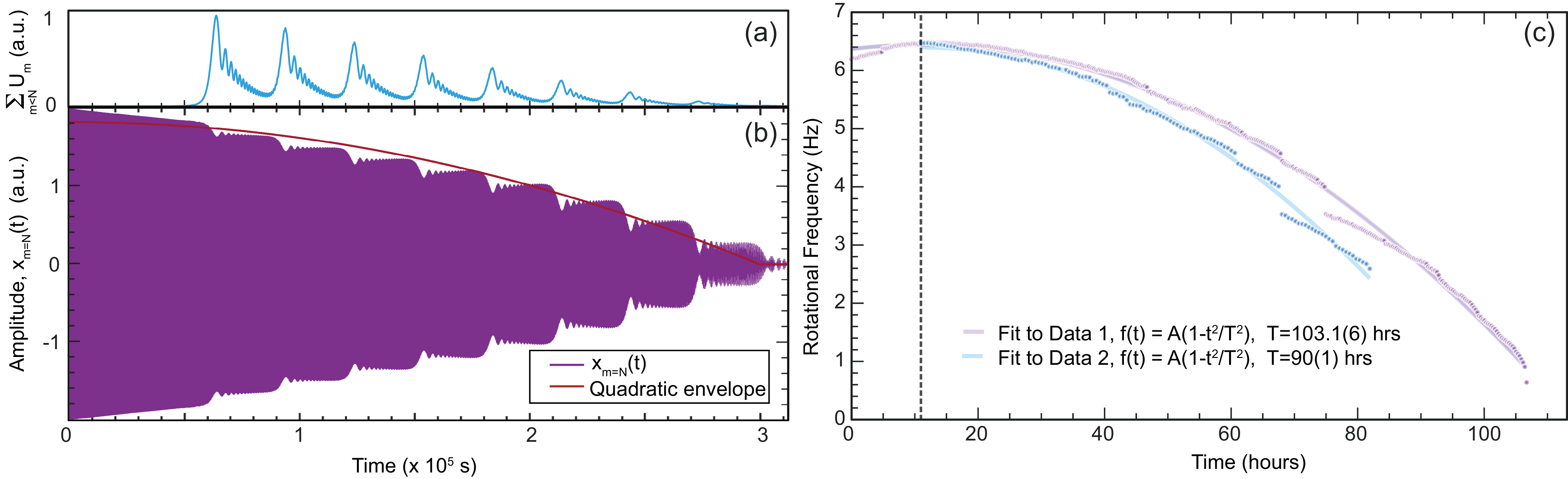}
    \caption{(a) Simulation of total energy, $\sum U_m$, accumulated in 9 lossy modes coupled to a drifting mode. Sharp bursts correspond to individual resonant crossings between the drifting mode and the fixed modes, demonstrating stepwise energy transfer out of the drifting mode. Here, a constant drifting rate and equally spaced lossy modes are considered. (b) Time evolution of the drifting-mode displacement $x_N(t)$ following an impulsive excitation. The red curve is the best-fit quadratic envelope, $A(t) = A_0\!\left[1 - (t/T)^2\right]$. For this simulation, it is assumed that the drifting mode dissipation rate is $\sim \SI{0.8}{nHz}$ while fixed modes have a dissipation rate of $\sim\SI{80}{\micro Hz}$. (c) Measured decay of the rotational frequency as a function of time is shown for two data sets. Unlike linear damping expected from eddy damping alone, the frequency exhibits a quadratic decay of the form $\omega(t) = \omega_0 (1 - (t/T)^2 ) $ indicative of amplitude-dependent energy loss mechanisms. We observe the same qualitative trend across multiple runs and different initial amplitudes. }\label{fig3}
\end{figure*}

\section{Mode Coupling and Anomalous Damping}

To first model the system, we consider a mechanical system with multiple coupled modes, where one mode (\emph{drifting mode}) has a frequency that evolves slowly over time and crosses several fixed-frequency modes. Let $x_i(t)$ denote the displacement of the $i$-th mode ($i = 1, \dots, N$). The equations of motion can be written as
\begin{equation}
    \ddot{\mathbf{x}} + \mathbf{C}\, \dot{\mathbf{x}} + \mathbf{K}(t)\, \mathbf{x} = 0,
\end{equation}
where $\mathbf{K}(t)$ is the time-dependent stiffness matrix encoding the instantaneous mode frequencies and coupling between modes, and $\mathbf{C}$ is the damping matrix.

At each time, the system has \emph{instantaneous normal modes}, obtained by diagonalizing $\mathbf{K}(t)$. When the drifting-mode frequency approaches that of a fixed mode, the coupling leads to hybridization of the two modes. In analogy with Landau-Zener theory~\cite{landau1932theorie, zener1932non} for quantum systems, the energy initially in the drifting mode partially transfers to the fixed mode. The fraction of energy transferred depends on the coupling strength and the rate of frequency change:
\begin{equation}
    P_{\rm transfer} \sim \exp\left(-\frac{\pi g^2}{\dot{\Delta}}\right),
\end{equation}
where $g$ is the coupling and $\dot{\Delta}$ is the instantaneous frequency difference rate.

As a result, the amplitude of the drifting mode exhibits sudden decreases at each crossing, while the energy in the fixed modes temporarily increases. When multiple modes are crossed, these energy exchanges create a non-monotonic decay envelope.

Even if the intrinsic damping of the drifting mode is small, the cumulative energy transfer to other modes acts as an \emph{effective damping}. Let $x_d(t)$ denote the displacement of the drifting mode. The effective damping equation can be approximated as
\begin{equation}
    \ddot{x}_d + \gamma_{\rm eff}(t) \dot{x}_d + \omega_d^2(t) x_d = 0,
\end{equation}
where $\gamma_{\rm eff}(t)$ is a time-dependent damping arising from mode coupling. When the intrinsic damping of the crossed modes is large (strongly lossy modes), energy transferred from the drifting mode is quickly dissipated, leading to an enhanced decay rate at the crossings.

In the regime where the drifting mode crosses multiple lossy modes in succession, the cumulative effect of these small, discrete energy transfers produces a decay of the amplitude that is approximately quadratic in time:
\begin{equation}
    x_d(t) \sim x_0 \left( 1 -  (t/\tau)^2 \right),
\end{equation}
for short-to-intermediate times, where $\tau$ is determined by the coupling strengths, damping of the fixed modes, and the frequency drift rate. 
Fig. \ref{fig3}(a)-(b) shows the numerical plot of a system of 9 lossy modes coupled to a drifting mode (total of 10 modes considered).  The energy transfer to other modes and a staircase decay of energy of the drifting mode can be seen when the drifting mode's frequency crosses one of the lossy modes' frequencies.

This quadratic-like decay emerges naturally from the \emph{distributed energy loss} mechanism: each mode crossing removes a small fraction of the energy, and the number of crossings increases linearly with time due to the drift of the mode frequency.

In our case, the drifting mode is the rotational frequency about the vertical axis and thus the amplitude decay is observed as decay of rotational frequency. Considering $\alpha(t)$ being the primary angular degree of freedom, $\{x_j(t)\}_{j=1}^N$ represent a set of higher frequency modes
with natural frequencies $\omega_j$.  The coupled system may be expressed as
\begin{align}
I \ddot{\alpha} + \gamma_\alpha \dot{\alpha}
+ \frac{\partial}{\partial \alpha} V_{\mathrm{eff}}(\alpha,\dot{\alpha})
&= \tau_{\mathrm{rad}}(t)
  - \sum_{j=1}^{N} k_j f_j(\dot{\alpha}) x_j, \label{eq:theta}\\
m_j \ddot{x}_j + \Gamma_j \dot{x}_j + m_j \omega_j^2 x_j
&= k_j g_j(\dot{\alpha}), \label{eq:xj}
\end{align}
where $f_j$ and $g_j$ encode nonlinear geometric couplings between the rotation
and other degrees of freedom.  The coefficients $k_j$ are typically
small but nonzero, and the internal modes satisfy $\Gamma_j \gg \gamma_\alpha$
due to enhanced dissipation.

When the instantaneous rotation frequency
$\Omega(t) = \dot{\alpha}(t)$ satisfies the resonance condition
\begin{equation}
\Omega(t) \approx \omega_j ,
\end{equation}
Eqs.~(\ref{eq:theta})--(\ref{eq:xj}) imply the emergence of a nonlinear
resonance window enabling enhanced energy transfer.  A matched-asymptotic
analysis shows that the effective damping of the slow mode acquires sharp
frequency-dependent corrections,
\begin{equation}
\gamma_{\mathrm{eff}}(\Omega)
=
\gamma_\alpha
+
\sum_{j}
\frac{k_j^{2}}{m_j \Gamma_j}
\frac{1}{1 + \left( \frac{\Omega - \omega_j}{\Gamma_j / 2} \right)^2 } ,
\end{equation}
which produces the observed discrete ``steps'' in the decay of $\Omega(t)$.
Each resonance crossing momentarily enhances dissipation, leading to a rapid
loss of energy until the mode detunes sufficiently.


\begin{figure*}
    \centering
\includegraphics[width=0.75\linewidth]{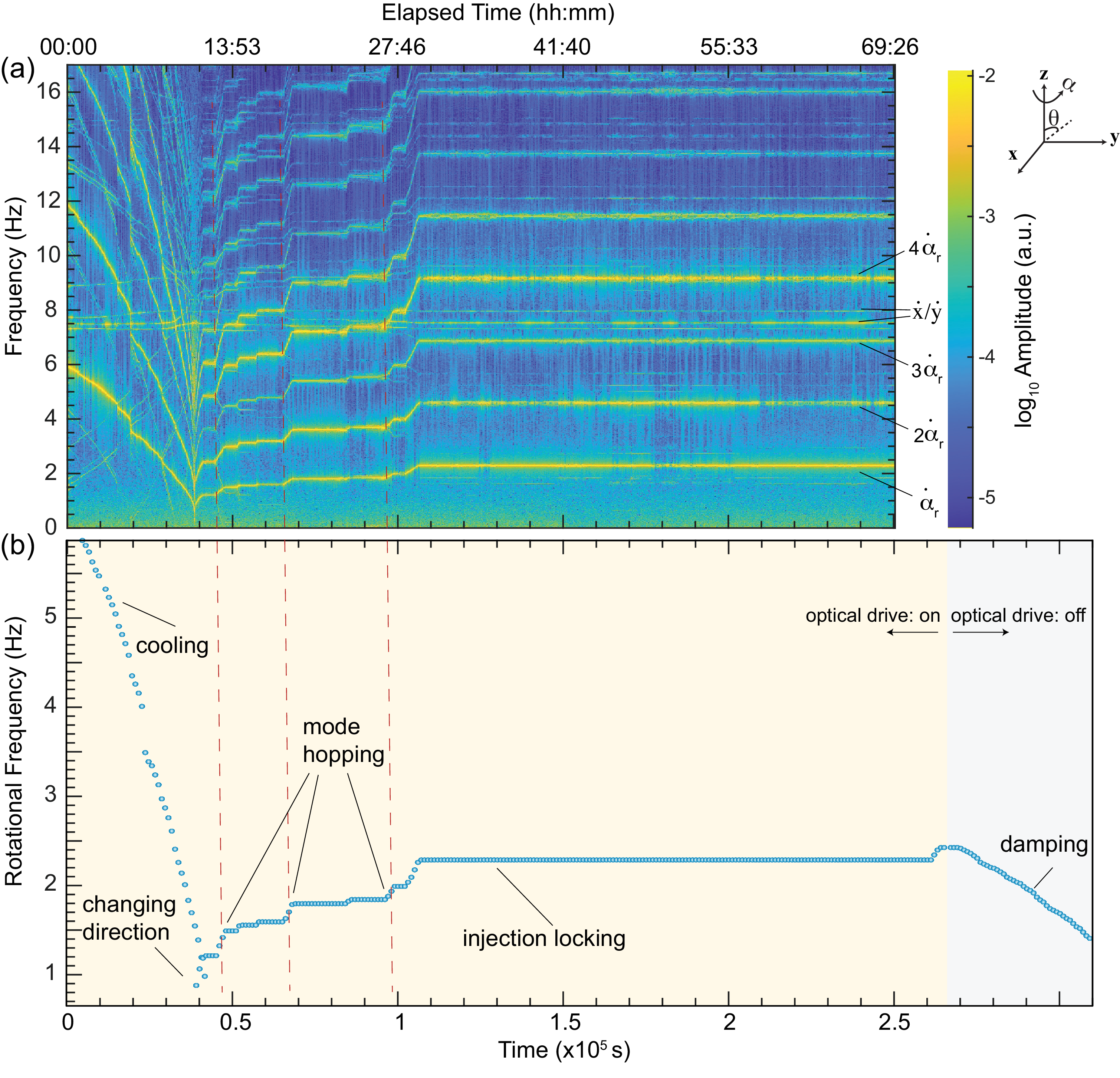}    
    \caption{(a) The evolution of the power spectral density (PSD) over time during a typical experiment is shown as measured by the top photo-detector. Initially, the maximum rotation frequency is reached using a strong side laser drive (100~mW for several seconds) and then a weak laser (1~mW) drive was applied from the top which is also used to probe the vibrations and calculate PSD. The weak drive applied an opposing torque decelerating the rotational motion. The frequency gradually decreases until the rotation direction reverses and starts to accelerate in the opposite direction. A striking feature is the self-locking of the rotation frequency: after the transient acceleration, the cube settles into a quasi-steady rotational state where the frequency drifts slowly over many hours.  (b) Rotational frequency measured using a side camera is plotted as a function of time replicating the dynamics observed in (a). At $t=2.6\times10^5$s the weak drive (top laser) is switched off and the dynamic of the particle was only recorded using the side camera. At this point the nominal quadratic decay is restored. The times at which mode crossing occurs are shown with dashed red lines.}
    \label{fig4}
\end{figure*}

To experimentally observe this effect, we spin the particle to more than 6~Hz or 360~RPM using the side laser drive and let the rotational energy decay while monitoring the rotational frequency using a high-speed camera from the top and the side. As seen in Fig.~\ref{fig3}(c), the decay of the rotational frequency shows jumps that overall can be modeled with a quadratic function. We attribute the quadratic damping to a weak transfer of angular momentum from the primary rotation to other rotational and translational modes of the cube, including small-angle rocking and translational motion within the magneto-gravitational potential. This cross-mode coupling is consistent with the multi-degree-of-freedom dynamics of large, nearly torque-free rigid bodies.
Fig.~\ref{fig3}(c) shows two data sets with similar long-time decay dynamics, although their initial behavior differs. In Data set 1, the rotational frequency decays monotonically, following the expected quadratic decay. In Data set 2, the drive laser was applied for a few seconds longer and at a slightly different location on the cube, resulting in the excitation of high-frequency translational and rocking modes. The drive was switched off just before the system entered a chaotic regime. Consequently, after the drive was turned off ($t=0$), the rotational frequency initially increased over the first few hours due to energy transfer from these high-frequency modes. Once the rotational frequency reached its maximum, the system recovered the quadratic decay behavior. 

\section{Stabilization and Emergent Order}

A striking phenomenon emerges when a weak counteracting radiation-pressure
torque $\tau_\mathrm{inj}$ is applied.  Instead of accelerating monotonically in
the reverse direction, the system evolves toward a persistent, highly stable
rotation rate that can persist for many hours. Such emergent order is a mechanical
analog of injection locking in nonlinear oscillators. This behavior, typically associated with optically trapped nanoparticles\cite{PhysRevLett.112.103603, brzobohaty2023synchronization} and torsional nanomechanical resonators\cite{Kuhn_2017}, is observed here for the first time in a millimetre-scale solid body.
Incorporating $\tau_\mathrm{inj}$ into Eq.~\ref{eq:theta} leads to an
Adler-type equation,
\begin{equation}
\dot{\Omega}
=
- \gamma_{\mathrm{eff}}(\Omega)\,\Omega
+ \frac{\tau_{\mathrm{inj}}}{I}, \label{eq:adler}
\end{equation}
but with a crucial distinction from optical injection locking:
$\gamma_{\mathrm{eff}}(\Omega)$ contains sharp nonlinear dispersion features
produced by the mode crossings.  A stable fixed point $\Omega^*$ is obtained
from
\begin{equation}
\gamma_{\mathrm{eff}}(\Omega^*) \, \Omega^*
=
\frac{\tau_{\mathrm{inj}}}{I},
\end{equation}
and because $\gamma_{\mathrm{eff}}(\Omega)$ contains multiple sharp peaks and
valleys, the fixed point can lie inside a plateau region that is dynamically
protected.  

To experimentally observe this effect, we start from a spinning cube at about 360~RPM and apply radiation pressure torque in the opposite direction of rotation. This is achieved using a weak (0.3~mW) laser light from the top. Due to the angled surface, reflection from the top surface at a particular laser-drive position exerts a very small radiation pressure torque, whose effect is only observable over hour-long timescales. Such weak excitation requires a low dissipation rate in the system. As seen in Fig.~\ref{fig4} (a)-(b), the opposing torque slowly stops rotation and reverses it after about 13 hrs. The enhanced damping induced by the light also undergoes mode hopping similar to Fig.\ref{fig3}. The particle eventually reverses direction and starts to spin while being weakly accelerated. Under this weak drive, unlike Fig.~\ref{fig2}(b), the rotational frequency does not increase continuously but in a staircase manner. The jump in rotational energy occurs when $x-y$ translational modes ($\sim 7.8$ Hz) crosses one of the higher order rotational modes. We note that the cube with eight corners exhibits not just the fundamental rotational mode but also higher order modes due to the distinct geometry of the object (e.g. eight corners) and the trap. We observe that after each jump in rotational frequency, when mode hopping happens, the frequency becomes increasingly more stable overtime. After another 14 hours the systems reaches a semi-steady state where the rotational frequency does not change for close to two days. When the drive is switched off the system undergoes typical dissipation as shown in Fig.\ref{fig4} (b). The observed effects of laser-induced damping (cooling), reversal, mode hopping, injection locking, and damping are also indicated in different parts of data in Fig.\ref{fig4} (b).      
The injection locking explains why the rotational frequency stabilizes rather than accelerating: the torque balances the locally minimized damping, yielding a long-lived attractor.



\section{Discussion and Conclusion}
As demonstrated here, levitation of macroscopic dielectric objects enables exceptionally low dissipation rates ($\gamma$), with rotational decay times extending to several days. We have previously shown\cite{sourki2025nonlinear} that a levitated object with mass $m \simeq 0.33$ mg can achieve acceleration sensitivities as low as $62~\mathrm{pg}/\sqrt{\mathrm{Hz}}$, outperforming state-of-the-art atomic interferometers. Combined with the absence of continuous external power during free evolution and the relatively compact experimental footprint, this platform presents a promising approach for precision inertial sensing. In particular, for acceleration measurements, sensitivity scales favorably with large values of $m/\gamma$, a regime naturally accessed by the system studied here.
Beyond low dissipation, increasing the sensing bandwidth requires access to higher resonance frequencies. In this work, we showed that radiation pressure can be used to controllably increase the rotational resonance frequency, up to a regime where nonlinear dynamics and chaotic behavior emerge. While such instabilities impose an upper limit on passive operation, they can be mitigated through active control to stabilize the system. In particular, cooling of high-frequency translational and rocking modes can suppress the onset of chaos, allowing rotational energy to increase to levels ultimately limited by the material’s structural strength, as observed in other systems \cite{schuck2018ultrafast}. Our observation that optical forces can both excite and damp motion indicates that dissipation engineering can be implemented entirely optically, without mechanical contact or cryogenic infrastructure. Cavity optomechanics is another way to significantly increase the resonance frequency\cite{PhysRevLett.111.183001} of such macroscopic objects. 

More broadly, our results demonstrate that even in the absence of bearings, contact, or friction, a rigid solid body can exhibit rich nonlinear dynamics traditionally associated with lasers, plasmas, and fluid systems. In this sense, the levitated cube constitutes a mechanical analogue of nonlinear optical oscillators: a minimal, tunable platform in which driven–dissipative nonlinear physics can be explored with high precision. Such a system opens new opportunities for applications in precision sensing \cite{leng2024measurement}, tests of fundamental physics \cite{tian2025constraints}, and the control of macroscopic motion at or near the quantum regime \cite{piotrowski2023simultaneous, vijayan2024cavity}. The intrinsic asymmetry of the cube plays a crucial role by enabling stable trapping and readout of all translational and rotational degrees of freedom. When used as a sensor, this allows simultaneous access to signals along multiple axes and orientations, enhancing both sensitivity and robustness.

A particularly powerful tool demonstrated here is injection-like locking of oscillations, which provides a means to control and stabilize the system in the presence of strong nonlinearity. The process  can be used to suppress frequency noise, stabilize the rotational phase, and enforce single-mode operation even near chaotic thresholds. For sensing applications, this enables narrowband amplification of weak signals, improved long-term stability, and enhanced readout fidelity.  
Together, these results establish levitated macroscopic rotors as a versatile platform for exploring nonlinear dynamics, engineered dissipation, and precision sensing, while providing new tools for stabilizing and controlling macroscopic motion in regimes previously considered inaccessible.

Our study demonstrates that nonlinear dynamical phenomena traditionally observed in complex, high-dimensional systems can arise in a low-dimensional system with a finite set of modes when the system parameters are carefully tuned.

\section*{Data availability}
The data supporting the findings of this study are available upon reasonable request.

\section*{Acknowledgments}
 This work was supported by DARPA Young Investigator Award, Grant No. D24AP00326-00.

%

\end{document}